\documentstyle[12pt,epsfig]{article}
\title{{\LARGE\bf Limit on a horizontal emittance in high energy muon colliders
due to synchrotron radiation.}
  \thanks{Talk at the Workshop Studies on Colliders and
    Collider Physics at the Highest Energies: Muon Colliders at 10 TeV
    to 100 TeV, 27 September - 1 October, 1999 Montauk, New York, USA,
    be published by the American Institute of Physics.  }  }

\author{Valery Telnov, \\
  {\small\it Budker Institute of Nuclear Physics, 630090 Novosibirsk, 
  Russia}\thanks{email:telnov@inp.nsk.su}}

\begin{document}
\newcommand{\EP}{\mbox{e$^+$}}
\newcommand{\EM}{\mbox{e$^-$}}
\newcommand{\EPEM}{\mbox{e$^+$e$^-$}}
\newcommand{\EMEM}{\mbox{e$^-$e$^-$}}
\newcommand{\GG}{\mbox{$\gamma\gamma$}}
\newcommand{\GE}{\mbox{$\gamma$e}}
\newcommand{\TEV}{\mbox{TeV}}
\newcommand{\GEV}{\mbox{GeV}}
\newcommand{\LGG}{\mbox{$L_{\gamma\gamma}$}}
\newcommand{\EV}{\mbox{eV}}
\newcommand{\CM}{\mbox{cm}}
\newcommand{\MM}{\mbox{mm}}
\newcommand{\NM}{\mbox{nm}}
\newcommand{\MKM}{\mbox{$\mu$m}}
\newcommand{\SEC}{\mbox{s}}
\newcommand{\CMS}{\mbox{cm$^{-2}$s$^{-1}$}}
\newcommand{\MRAD}{\mbox{mrad}}
\newcommand{\IND}{\hspace*{\parindent}}
\newcommand{\E}{\mbox{$\epsilon$}}
\newcommand{\EN}{\mbox{$\epsilon_n$}}
\newcommand{\EI}{\mbox{$\epsilon_i$}}
\newcommand{\ENI}{\mbox{$\epsilon_{ni}$}}
\newcommand{\ENX}{\mbox{$\epsilon_{nx}$}}
\newcommand{\ENY}{\mbox{$\epsilon_{ny}$}}
\newcommand{\EX}{\mbox{$\epsilon_x$}}
\newcommand{\EY}{\mbox{$\epsilon_y$}}
\newcommand{\BI}{\mbox{$\beta_i$}}
\newcommand{\BX}{\mbox{$\beta_x$}}
\newcommand{\BY}{\mbox{$\beta_y$}}
\newcommand{\SX}{\mbox{$\sigma_x$}}
\newcommand{\SY}{\mbox{$\sigma_y$}}
\newcommand{\SZ}{\mbox{$\sigma_z$}}
\newcommand{\SI}{\mbox{$\sigma_i$}}
\newcommand{\SIP}{\mbox{$\sigma_i^{\prime}$}}
\date{}
\maketitle
\begin{abstract}
It is shown that at a 100 TeV muon collider the synchrotron radiation
in the ring will determine the minimum horizontal emittance. 
\end{abstract}

  In all \EPEM\ storage ring the horizontal emittance is determined by
quantum nature of the synchrotron radiation. Emission of photons in
regions with non-zero dispersion leads to  growth of the horizontal
emittance while the average energy loss leads to the cooling. As
the result, some equilibrium emittance is reached. This is well known
and corresponding formulae can be found elsewhere~\cite{Wid}.

 At muon colliders, the synchrotron radiation power is much smaller
than that at \EPEM\ storage rings due to higher particle mass: $P
\propto (EB)^2/m^4$; however this suppression is compensated by the
much higher energy and magnetic field.

  In general,  emittance in a storage ring depends on  time in
  the following way
\begin{equation}
\EX\ = \epsilon_{D}(1-e^{-t/\tau}) + \epsilon_0 e^{-t/\tau},
\end{equation}
where $\epsilon_0$ is the initial emittance, $\epsilon_D$ is the
equilibrium emittance, $\tau$ is the damping time. One can check (see
also the B.King's table) that for the 2E=100 TeV muon collider
$\gamma_{\mu}\tau_{\mu}\approx \tau_D \approx$ 1 sec, so the emittance
 will be close to the equilibrium one.

The equilibrium {\it normalized} ($\epsilon_n=\gamma\epsilon$) 
horizontal emittance in a damping ring~\cite{Wid}
\begin{equation}
\epsilon_D = \frac{55\hbar
}{32\sqrt{3}mc}\frac{\gamma^3}{J_x}\frac{\langle H \rangle}{\rho},
\end{equation}
where $J_x\approx 1$ is the damping partition number, $\langle H
\rangle \propto \beta^3/\rho^2$ is the average value of the
``H-function'' characterizing a storage ring magnetic structure. One
can see that for fixed energy and magnetic field $\ENX\ \propto
1/m.^4$

For an optimized FODO ring structure of the muon collider the minimum
normalized horizontal emittance due to synchrotron radiation~\cite{Wid}
\begin{equation}
\epsilon_D \sim 100
\frac{E^3[\GEV]}{N^3}\left(\frac{m_e}{m_{\mu}}\right)^4\;\;\mbox{cm rad},
\end{equation}
where $N$ is the number of bending magnets in the ring. Extrapolating
the number of magnets in the current designs to the 100 TeV energy as
$N \propto \sqrt{E}$ (the usual energy dependence of the
$\beta$-function) one can get $N\sim 1000$ for $2E=100$ TeV. For this
structure we obtain the normalized horizontal emittance at $t=\tau_D$
\begin{equation}
\ENX\ \approx 0.63 \epsilon_D \approx 5\times 10^{-3}\; \mbox{cm rad}.
\end{equation}
For comparison, in the B.King's table 
$\ENX\ = 8.7\times 10^{-4}$ cm rad for ``evolutionary'' extrapolation and 
$\ENX\ = 2.1\times 10^{-5}$ cm rad for ``ultra-cold beams'', 
which is smaller than above result by  factors of 5 and 250, respectively.

  This example shows that the considered effect is important for high
energy muon colliders and this problem needs more accurate
consideration. One should consider the more realistic case of the isochronous
ring (which is necessary for muon colliders), including W-shielding which
reduces the quad strength (and correspondingly $N$).

\vspace*{0.5cm}


\begin{thebibliography}{9}
\bibitem{Wid} H.Wiedemann, Particle Accelerator Physics, v.1, Springer-Verlag.
\end{thebibliography}
\end{document}